\documentclass[aps,prb,preprint]{revtex4-1}

\usepackage{graphicx}

\usepackage[pdftex,colorlinks=true,hidelinks]{hyperref}
\hypersetup{
pdftitle={Increasing the Efficiency of a Thermionic Engine Using a Negative Electron Affinity Collector},
pdfauthor={Joshua Ryan Smith (joshua.smith133.ctr@mail.mil)},
pdfsubject={},
pdfkeywords={},
pdfcreator={pdfTeX}
}
\begin{document}

\title{Increasing the Efficiency of a Thermionic Engine Using a Negative Electron Affinity Collector}
\author{Joshua Ryan Smith}
\email[]{joshua.smith133.ctr@mail.mil}
\affiliation{US Army Research Laboratory}

\date{\today}

\begin{abstract}
Most attention to improving vacuum thermionic energy conversion device (TEC) technology has been on improving electron emission with little attention to collector optimization. A model was developed to characterize the output characteristics of a TEC where the collector features negative electron affinity (NEA). According to the model, there are certain conditions for which the space charge limitation can be reduced or eliminated. The model is applied to devices comprised of materials reported in the literature, and predictions of output power and efficiency are made, targeting the sub-1000K hot-side regime. By slightly lowering the collector barrier height, an output power of around $1kW$, at $\geq 20\%$ efficiency for a reasonably sized device ($\sim 0.1m^{2}$ emission area) can be achieved.
\end{abstract}


\maketitle

\section{Introduction}
Producing more energy at high efficiency is a constant challenge facing humanity. Thermionic energy conversion has a role to play due to its many advantages: the design is simple, there are no moving mechanical parts, and there is no efficiency-reducing direct conduction of heat across the device \cite{978-0-309-08686-8, 10.1063/1.1702392}.

Despite the advantages of thermionic engines, there have been technical challenges that have prevented adoption on a large scale. The challenges have been dealing with the high temperatures required to achieve a reasonably good efficiency and having to overcome the effects of a space charged limited mode.

The first challenge is to create a TEC which operates at a reasonably low temperature, but still outputs acceptable power at high efficiency. In order to get the same electron emission from a material at lower temperature, one must either find a way to lower the thermionic barrier of that material, or one has to find a different material with a lower barrier.

Na\"{i}vely, one might think that lowering the emission barrier of the emitter at all costs is the way to improve electron emission and therefore TEC output power density. It is true that electron emission dramatically increases by lowering the barrier, but at the device level output power will not necessarily increase. One reason is that more emitter output current results in a greater negative space charge effect. Therefore emitter barrier lowering quickly hits a point of diminishing returns due to the self-limiting nature of the negative space charge effect. The second general issue stems from the fact that output power depends on the product of output current \emph{and} the operating voltage. Negative space charge issues aside, the optimal operating voltage of the device is nominally the difference in barrier heights of the two electrodes. A strategy of lowering the emitter barrier without consideration of the collector will also quickly reach a limit of effectiveness as the value of the emitter barrier approaches that of the collector.

The second challenge is the so-called negative space charge effect. The output current of a TEC is limited by the negative space charge effect because electrons traversing the interelectrode space create a negative charge barrier which blocks lower energy electrons from reaching the collector. Decreasing the interelctrode spacing is one approach to space charge mitigation.

Generally speaking, the emitter barrier should be greater than the collector barrier. The emitter barrier should be low and its Richardson's constant should be high such that enough current is available to reach the desired output power. The electrodes should be close enough together to mitigate the negative space charge effect, bearing in mind that decreased spacing increases the engineering difficulty of device fabrication.

Recent work has focused on improving the emission characteristics of the emitter electrode. Nemanich, Koeck, and collaborators have reported favorable emission parameters from nitrogen \cite{10.1016/j.diamond.2008.11.023} and phosphorus doped diamond \cite{10.1016/j.diamond.2009.01.024}. Further improvements in electron emission occur when a sample is irradiated with light via a phenomenon called photon-enhanced thermionic emission (PETE). Schwede et. al. first observed PETE from GaN \cite{10.1038/nmat2814} and Sun et. al. have also observed it from diamond \cite{10.1063/1.3658638}.

Regarding space charge mitigation, Hatsopoulous and Kaye demonstrated at considerable difficulty that a close-spaced configuration mitigates space charge \cite{10.1063/1.1723373}. More recently, Lee and collaborators have applied MEMS fabrication techniques to decrease the interelectrode distance \cite{lee}. Smith et. al. have predicted that an emitter which features a negative electron affinity, such as hydrogen terminated diamond, could  mitigate or eliminate the negative space charge effect \cite{10.1103/PhysRevB.76.245327,10.1116/1.3125282}.

Improvement of the emission properties are welcome advances in the technology, but many materials already exist with quite good emission at low temperatures: scandate electrodes, BaO, CsO, and other oxide materials \cite{10.1016/S0169-4332(96)00698-8} all have exceptional electron emission properties.

Little investigation is being conducted on improving the collector electrode technology. In this paper a model is developed which considers the effect of a negative electron affinity collector on electron transport through a vacuum thermionic engine. It is shown that the negative electron affinity reduces, or in some cases eliminates the negative space charge effect. The model is applied to a TEC which is comprised of materials reported in the literature and could conceivably be constructed today. Finally, it is shown that of all the device parameters, the output power density and efficiency can be most effectively optimized by lowering the collector barrier. The model predicts that a reasonably sized device ($< 1m^{2}$ emission area) can produce around $1kW$ at $\geq 20\%$ efficiency at a hot-side temperature of $\approx 1000K$.

\section{Electron transport}
\begin{figure}
\includegraphics{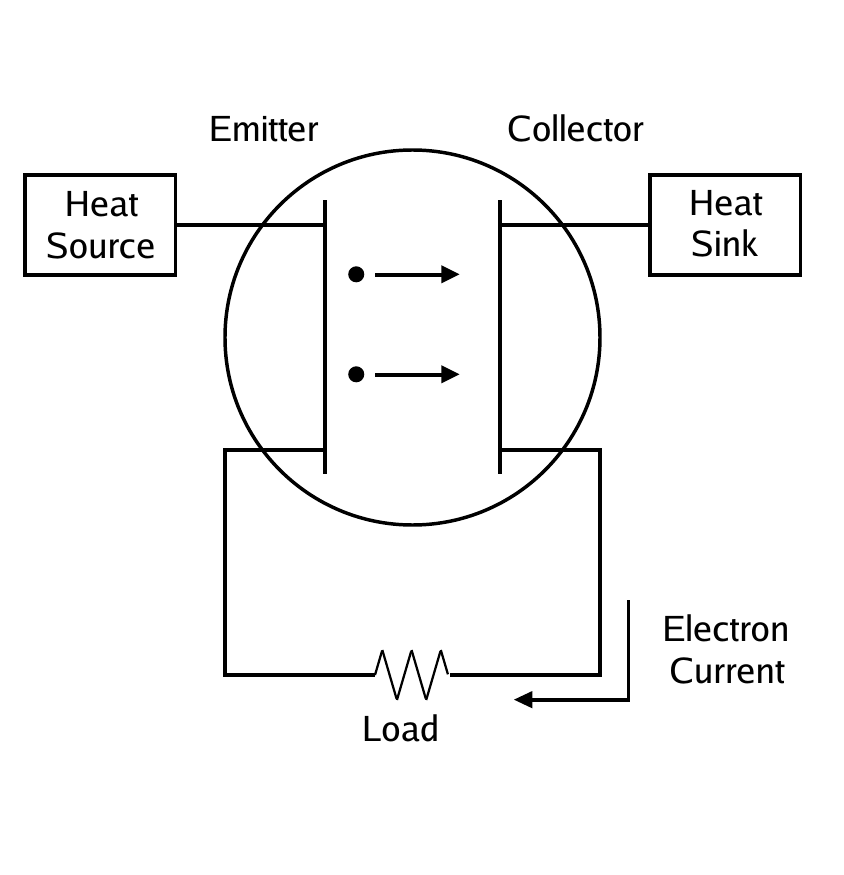}
\caption{General schematic of a vacuum TEC. The emitter and collector electrodes are enclosed in a vacuum container and separated by some distance. The emitter is in thermal contact with a thermal reservoir at a higher temperature, and the collector is in thermal contact with a thermal reservoir at a lower temperature. Electrons are thermionically emitted from the emitter, travel across the interelectrode space, and are absorbed by the collector. The electrons travel from the collector through an electrical lead, through an external load where work is done, and back to the emitter to complete the circuit.}
\label{fig:02}
\end{figure}

A thermionic engine (aka thermionic energy conversion device or TEC) is a vacuum device that converts heat directly to electrical work. It is a heat engine and can be analyzed as such. A schematic diagram is shown in Fig. \ref{fig:02}.

\begin{figure}
\includegraphics{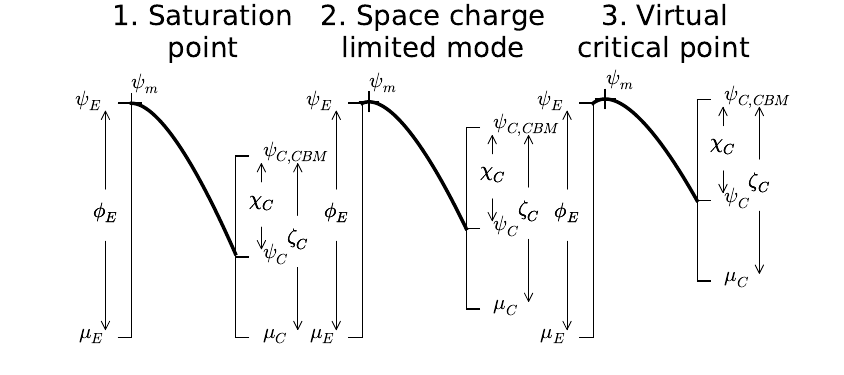}
\caption{Various conditions of electron motive in a TEC where the collector features NEA. The emitter is depicted on the left on each subplot, and the collector on the right. The maximum motive is indicated by a '+' symbol.}
\label{fig:03}
\end{figure}

Electron transport is most easily understood by use of the electron motive and electron motive diagram; the motive diagram is similar to the band diagram in a solid-state device. Examples of a motive diagram in the case where the emitter electrode has PEA and the collector has NEA are given in Fig. \ref{fig:03}. Subscripts, E, and, C, denote emitter and collector, respectively. The electrodes are separated by a distance, $d$. The motive at any point is denoted by $\psi$, and the vacuum level at either surface is denoted by $\psi$ appropriately subscripted (e.g. in the case of the emitter: $\psi_{E}$). The maximum motive is denoted by $\psi_{m}$. The quantity $eV$ represents the potential of the collector relative to the emitter, $V$ is referred to as the output voltage and $e$ is the fundamental electronic charge. The output voltage is physically determined by the output current and the external electrical load of the device. For the case of PEA, $\phi$ denotes the thermionic barrier (sometimes referred to as work function), and in the case of NEA the barrier is denoted by $\zeta$ which is the difference between the conduction band minimum and the Fermi energy. The value of NEA is denoted by $\chi$. The quantity $\psi_{C,CBM}$ is the conduction band minimum of the collector, and $\mu$ denotes the Fermi energy.

Electrons inside the emitter with energy greater than $\psi_{E}$ are thermionically emitted into the vacuum. The current density of the emitted electrons is given by Richardson's equation.

\begin{equation} \label{eq:00}
J = AT^{2}\exp\left( -\frac{\phi}{kT} \right)
\end{equation}

\noindent where $J$ is the emission current [$A cm^{-2}$], $A$ is Richardson's constant, [$A cm^{-2} K^{-2}$], $T$ is the temperature [$K$], $\phi$ is the barrier height or work function [$eV$], and $k$ is Boltzmann's constant [$8.62 eV K^{-1}$].

Since the device is evacuated, electrons traveling across the interelectrode space represent a net negative charge. For some values of voltage, this negative space charge can cause the  motive in the interelectrode space to be greater than the emitter barrier, presenting an additional space charge barrier to the thermionic electrons. This negative space charge effect decimates the output current for most devices and is a significant challenge in creating a viable thermionic device. 

\section{Theory}
Electron transport in the face of the negative space charge effect is modeled using a Vlaslov-Poisson system, following the success of this approach in the past \cite{10.1103/PhysRev.21.419, 10.1063/1.1735850}. Consider a TEC with a PEA emitter and a NEA collector. We would expect this device should experience the negative space charge effect under most typical operating conditions, but we assume the collector is sufficiently cool so that back emission is negligible. Electrons arriving at the collector with energy greater than $\psi_{C,CBM}$ are assumed to be absorbed by the collector.

As the output voltage changes, this device will pass through several unique modes of electron transport. The accelerating mode is the condition where the maximum motive occurs at the emitter electrode. The saturation point is the greatest value of output voltage such that the maximum motive occurs just outside the emitter; it represents a bound on the accelerating mode and is depicted in Fig. \ref{fig:03}.1.

As the output voltage increases, the maximum motive exists somewhere within the interelectrode space. The maximum motive barrier reduces the current traveling across the device, and so this set of voltages is known as the space charge limited mode. The motive diagram for a typical point in the space charge limited mode is depicted in Fig. \ref{fig:03}.2. The virtual critical point occurs when the maximum motive and collector conduction band minimum coincide at the same height. This condition is shown in Fig. \ref{fig:03}.3.

As the voltage increases, the conduction band minimum of the collector limits the electron current entering the collector: electrons with energy less than the value of $\psi_{C,CBM}$ cannot enter the collector and are scattered back to the emitter.

At some value of voltage, the maximum motive occurs immediately outside the collector. This situation is referred to as the critical point. For voltages greater than the critical point voltage, the TEC is in the retarding mode.

Between the critical point and the virtual saturation point, the analysis of Langmuir \cite{10.1103/PhysRev.21.419} exactly models the electron transport. For output voltage above the virtual saturation point, the electron transport differs from Langmuir but the current is ultimately limited by the position of $\psi_{C,CBM}$. Therefore the output current is expressed by Eq. \ref{eq:11} which is unaffected by the precise details of the motive in the interelectrode space.

\begin{equation} \label{eq:11}
J = A T_{E}^{2} \exp \left( -\frac{\zeta_{C} + eV}{k T_{E}} \right)
\end{equation}

Notation from Langmuir's analysis is used in the following derivations and the highlights are listed here. In the space charge limited mode, Langmuir converts the motive and position to the dimensionless quantities; \mbox{$\gamma \equiv \frac{\psi_{m} - \psi}{k T_{E}}$} is the dimensionless motive and \mbox{$\xi = (x - x_{m}) \left(\frac{2 \pi m_{e} e^{2}}{\epsilon_{0}^{2} k^{3}}\right)^{1/4} \frac{J^{1/2}}{T_{E}^{3/4}}$} is the dimensionless position. The quantity, $x_{m}$, is the position where $\psi_{m}$ occurs. It is worth noting that the appearance of $J$ in the expression for $\xi$ is an indication that we are self-consistantly solving the Vlaslov and Poisson equations. These dimensionless quantities allow Langmuir to write Poisson's equation in a universal form and thus numerically calculate the solution. Hatsopoulous and Gyftopoulous give a clear exposition of an algorithm to determine the output voltage given a value of space-charge limited current density \cite{0-262-08059-1}. Thus, the output current density vs. output voltage can be determined for a set of operating parameters.

\subsection{Saturation Point}
In this case the output current is given by the saturation current of the emitter, and the saturation point dimensionless distance at the collector is given by

\begin{equation} \label{eq:01}
\xi_{CS} = d \left(\frac{2 \pi m_{e} e^{2}}{\epsilon_{0}^{2} k^{3}}\right)^{1/4} \frac{J_{ES}^{1/2}}{T_{E}^{3/4}}
\end{equation}

\noindent This value corresponds to a value of $\gamma_{CS}$ according to Langmuir's solution to the dimensionless Poisson's equation. From Fig. \ref{fig:03}.1 and the definition of $\gamma$, 

\begin{equation} \label{eq:02}
eV_{S} = \phi_{E} + \chi_{C} - \zeta_{C} - \gamma_{CS} k T_{E}
\end{equation}

\subsection{Virtual Critical Point}
By the definition of the virtual critical point, and referencing Fig. \ref{fig:03}.3,

\begin{equation} \label{eq:03}
\chi_{C} = \psi_{m} - \psi_{CVR}
\end{equation}

\noindent The subscript \emph{CVR} indicates \emph{C}ollector \emph{V}irtual c\emph{R}itical point. Using the definition of $\gamma$ and substituting,

\begin{equation} \label{eq:05}
\gamma_{CVR} = \frac{\chi_{C}}{k T_{E}}
\end{equation}

\noindent Using the value of $\gamma_{CVR}$, one can use the dimensionless solution to Poisson's equation to determine $\xi_{CVR}$. From the definition of $\xi$,

\begin{equation} \label{eq:12}
\xi_{EVR} = \xi_{CVR} - d \left(\frac{2 \pi m_{e} e^{2}}{\epsilon_{0}^{2} k^{3}}\right)^{1/4} \frac{J_{VR}^{1/2}}{T_{E}^{3/4}}
\end{equation}

Using the dimensionless Langmuir solution, $\gamma_{EVR}$ can be determined from $\xi_{EVR}$. From the virtual critical point motive diagram and the definition of $\gamma$,

\begin{equation} \label{eq:08}
e V_{VR} = \phi_{E} - \zeta_{C} + k T_{E} \gamma_{EVR}
\end{equation}

\subsection{Space Charge Limited Mode}
The method of calculating the output current characteristic in the space charge limited mode is nearly identical to the Langmuir case, one difference being the space charge mode extends from the saturation point only to the virtual critical point when the collector exhibits NEA. Hatsopoulous and Gyftopoulous's algorithm to find the output voltage given a value of output current density can be adapted to this case.

\begin{enumerate}
\item Given $J$, calculate $\gamma_{E}$ using Eq. \ref{eq:23}.
\item Calculate $\xi_{E}$ using Langmuir's solution to the dimensionless Poisson's equation.
\item Compute $\xi_{C}$ from Eq. \ref{eq:21}.
\item Again using Langmuir's solution to the dimensionless Poisson's equation, calculate $\gamma_{C}$ from $\xi_{C}$.
\item The output voltage is given by Eq. \ref{eq:22}.
\end{enumerate}

\begin{equation} \label{eq:23}
\gamma_{E} = ln \left( \frac{J_{ES}}{J} \right)
\end{equation}

\begin{equation} \label{eq:21}
\xi_{C} = \xi_{E} + d \left(\frac{2 \pi m_{e} e^{2}}{\epsilon_{0}^{2} k^{3}}\right)^{1/4} \frac{J_{VR}^{1/2}}{T_{E}^{3/4}}
\end{equation}

\begin{equation} \label{eq:22}
eV = \phi_{E} + k T_{E} \gamma_{E} - (\zeta_{C} - \chi_{C} + k T_{E} \gamma_{C})
\end{equation}

\section{Analysis}
\subsection{The model applied to a plausible TEC}
\begin{figure}
\includegraphics{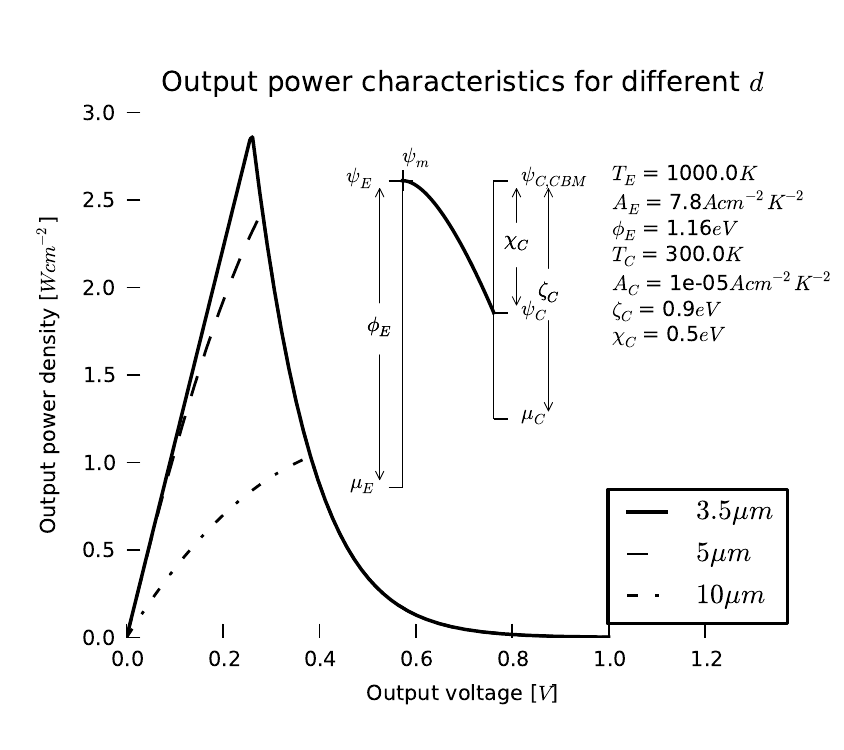}
\caption{Output power characteristic showing the reduction to zero of the space charge limited mode by decreasing the interelectrode spacing. Inset: Motive diagram at the condition where the saturation and virtual critical point coincide.}
\label{fig:06}
\end{figure}

The model can be used to predict the performance of a TEC created from materials reported in the literature. Consider a TEC constructed from a scandate electrode described by G\"{a}rtner \emph{et.al.} \cite{10.1016/S0169-4332(96)00698-8} as the emitter, and a phosphorous doped diamond electrode described by Koeck \emph{et.al.} \cite{10.1016/j.diamond.2009.01.024} as the collector. The emitter has a barrier of $1.16eV$ and a Richardson's constant of $7.8A cm^{-2} K^{-2}$, while the collector has a barrier of $0.9eV$ and a Richardson's constant of $1 \times 10^{-5}A cm^{-2} K^{-2}$. Efficiency of the device can be calculated by considering the electronic heat transport and the Stefan-Boltzmann heat transport; unfortunately no Stefan-Boltzmann emissivity data is quoted for either material. The scandate material is usually applied as a thin coating to a refractory metal such as tungsten, and the diamond is CVD deposited as a thin film on molybdenum. Thus, a mid-range emissivity value of 0.5 is assumed for both electrodes. No value is quoted for the NEA of the phosphorus doped diamond collector. According to the literature, the magnitude could be as large as 1eV \cite{10.1103/PhysRevB.50.5803, 10.1103/PhysRevLett.81.429}, so a conservative value of $0.5eV$ is chosen. The interelectrode spacing is chosen to be $10 \mu m$. Finally, assume the emitter is held at a temperature of $1000K$ while the collector is at $300K$. At such low temperature, the back thermionic emission from the collector is negligible, consistent with the model. The output power characteristic of these parameters as calculated by the model is given by the dash-dot line in Fig. \ref{fig:06}. The  maximum output power density is $1.03 W cm^{-2}$ at an efficiency of $17.8\%$ and occurring at an output voltage of $0.38V$. Thus, the emission area of the device would be $0.1 m^{2}$ to produce $1kW$ output power. The output power density and emission area of this device are reasonable, but the efficiency falls short of the 20\% target.

\begin{figure}
\includegraphics{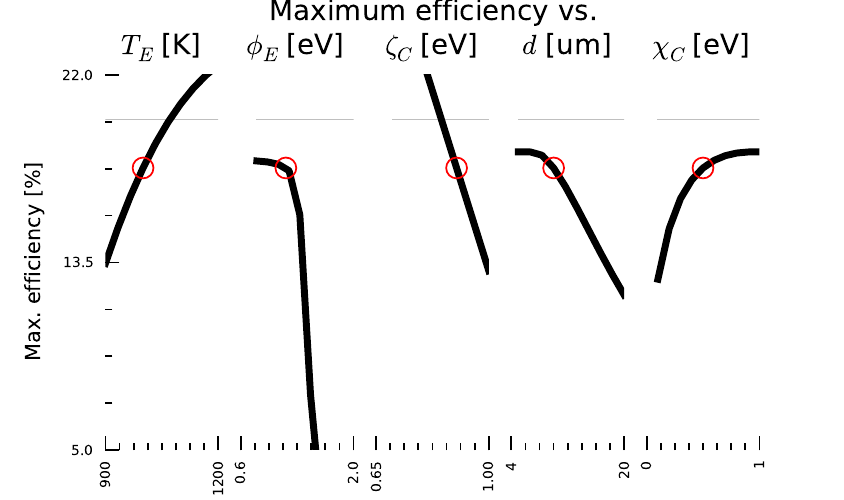}
\caption{First row: maximum efficiency vs. various TEC parameters. Second row: corresponding output power density. Third row: corresponding output voltage. The red circles indicate the common starting set of parameters.}
\label{fig:13}
\end{figure}

In the physical world, materials parameters such as Richardson's constant, emissivity, and barrier height are usually coupled. The computer model is not constrained by this coupling and so individual parameters can be adjusted to quickly see which parameters have the biggest effect on device performance. This simulation is valuable to build intuition about how this system behaves since the system does not have a closed-form solution. Using the device just described as a starting point, individual input parameters are adjusted (keeping all others fixed) to calculate the maximum efficiency. Fig. \ref{fig:13} depicts the results of such a calculation as an array of plots. The of plots depict the maximum efficiency vs. a particular input parameter; the plots from left to right show the effect of adjusting the emitter temperature, emitter barrier, collector barrier, collector NEA, and interelectrode spacing, respectively. The red circles indicate the value of the common starting set of parameters.

The first plot shows that the efficiency increases as emitter temperature increases, crossing 20\% efficiency at $T_{E} = 1070K$. This result is not surprising because increasing the emitter temperature increases the number of electrons in excited states and thus more electrons are available to cross the device.

The second plot shows that efficiency increases rapidly as the emitter barrier decreases, then levels off. If the emitter barrier is high, few electrons escape. The output current is small, and therefore the negative space charge effect is negligible. Additionally, since the output current is small, both the output power and efficiency will be small as well. As the emitter barrier decreases, the output current increases but so does the negative space charge barrier. At some point, the emitter barrier is low enough that the current is limited mostly by the space charge and not the emitter barrier; further lowering the emitter barrier has a diminishing effect on the output current. The space charge limitation on current limits the output power and thus efficiency. Lowering the emitter barrier is akin to integrating over a larger interval of the thermal distribution of the electrons of the emitter, in contrast to increasing the emitter temperature which increases the scale of the distribution.

The third plot shows that the maximum efficiency increases as the collector barrier is lowered, crossing 20\% at $\zeta_{C} = 0.85eV$. This condition yields an output power density of $1.17W cm^{-2}$ (corresponding to an emission area of $0.086 m^{2}$ for $1kW$ total output) at an output voltage of $0.43V$. Examining the motive diagram is the easiest way to understand the relationship between collector barrier and maximum efficiency. Consider that the efficiency is calculated as the output power divided by the rate of heat input, and the output power is the product of the output current and voltage. In the motive diagrams depicted in Fig \ref{fig:03}, the  motive due to the negative space charge effect is calculated using the vacuum levels of both electrodes as boundary conditions. On the collector side, the vacuum level is determined relative to the emitter Fermi energy by the output voltage (times the fundamental charge), the collector barrier, and the collector NEA. For a smaller value of collector barrier, the system would have the same boundary conditions if the output voltage was increased (assuming constant value of NEA). This tradeoff does not affect the output current of the device, but the output power would increase as a result of the increased voltage. Moreover, the efficiency would increase as well. This relationship between collector barrier and efficiency is not unique to the case where the collector features NEA -- a similar relationship holds in a traditional TEC.

Strictly speaking, collector barrier reduction has a limit of effectiveness because at some point the collector will experience appreciable back emission. Back emission is negligible over the entire range of collector barrier height values shown in Fig. \ref{fig:13}. The back current attains a value of $1e-7 A$ at $\zeta_{C} = 0.65eV$, which is many orders of magnitude smaller than the output current.

The fourth and fifth plots show that efficiency tapers off as the value of NEA increases and the interelectrode distance decreases, respectively. These results are two manifestations of the same phenomenon having to do with space charge mitigation as explained in the following section. Without going into detail, when the collector exhibits NEA, there exists a set of device parameters such that the negative space charge effect is completely eliminated. Once the space charge is eliminated, the device operates at its ideal performance; decreasing the interelectrode distance or increasing the NEA has no further effect on the performance of the device, and therefore the efficiency does not improve further.

\subsection{Reduction/elimination of the space charge mode via collector NEA}
In some cases, the model of the TEC under consideration will experience no space charge limited mode. There exists a volume in parameter space such that for points inside the volume and on the surface bounding the volume, there is no range of voltages for which the TEC passes through a space charge limited mode. These are referred to as the ideal volume and ideal surface. Before the derivation of the proof, consider the following example which illustrates this phenomenon. Several output current characteristics for the example device are depicted in Fig. \ref{fig:06}, each calculated for a different interelectrode spacing, $d$. As $d$ is reduced, the negative space charge effect is also reduced since electrons spend less time in the interelectrode space. In Fig. \ref{fig:06}, this reduction in negative space charge effect is manifest in that the saturation point and virtual critical point move closer to one another as $d$ decreases. For some value of $d$, called $d_{ideal}$, the critical and virtual saturation points coincide; this combination of parameters lies on the aforementioned ideal surface in parameter space. At this point, the negative space charge effect is completely eliminated and there is no range of voltages for which the device passes through a space charge limited mode. For values of $d$ less than $d_{ideal}$, the performance of the device does not improve, hence the leveling off of maximum efficiency in the fifth plot of Fig. \ref{fig:13}.

It is well known that the negative space charge effect can be mitigated by reducing the interelectrode spacing of a vacuum thermionic engine; space charge mitigation in a device with an NEA collector is similar to space charge mitigation in a non-NEA device. The difference is that in the non-NEA device, full space charge elimination occurs in the limit of $d \rightarrow 0$, while in a NEA collector device full space charge elimination can occur at a finite distance. Lee et.al. have shown for very small values of interelectrode spacing the efficiency of a TEC degrades due to near-field radiative coupling \cite{10.1063/1.4707379}. According to the model described in this paper, a TEC featuring a negative electron affinity collector can achieve ideal operation with an interelectrode spacing sufficiently large to avoid near-field effects. The initial device considered in this report will be on the ideal surface in parameter space if the interelectrode distance is $3.89 \mu m$.

The inset of Fig. \ref{fig:06} depicts the general motive diagram corresponding to a point on the ideal surface. To derive this condition, consider a TEC with parameters on the ideal surface: it is equivalent to say the saturation point and virtual critical point of that TEC coincide. We first compute the quantities at the saturation point. From the figure, $\gamma_{CS}$ becomes

\begin{equation} \label{eq:13}
\gamma_{CS} = \frac{\chi_{C}}{k T_{E}}
\end{equation}

\noindent So Eq. \ref{eq:02} reduces to

\begin{equation} \label{eq:14}
eV_{S} = \phi_{E} - \zeta_{C}
\end{equation}

\noindent Note also that the output current density is equal to the emitter saturation current density.

Next we compute quantities of the virtual critical point. From the figure it is clear that the output current density must  also be equal to the emitter saturation current density. Moreover, $\xi_{EVR}$ equals zero and therefore $\gamma_{EVR}$ does as well. Eq. \ref{eq:08} becomes

\begin{equation} \label{eq:18}
e V_{VR} = \phi_{E} - \zeta_{C}
\end{equation}

\noindent The output voltage and output current density of both the saturation and virtual critical points are identical and therefore coincide.

Another consequence of the fact that $\xi_{EVR} = 0$ and $J_{VR} = J_{ES}$ is that Eq. \ref{eq:12} can be rewritten as

\begin{equation} \label{eq:19}
\xi_{CVR} = d \left(\frac{2 \pi m_{e} e^{2}}{\epsilon_{0}^{2} k^{3}}\right)^{1/4} \frac{J_{ES}^{1/2}}{T_{E}^{3/4}}
\end{equation}

Note that the left hand side of Eq. \ref{eq:19} depends on the value of $\gamma_{CVR}$ via Langmuir's solution to the dimensionless Poisson's equation. If the above condition is met, the saturation point coincides with the critical point and the TEC is on the ideal surface in parameter space.

Notice also that there are device parameters on which the left hand side of Eq. \ref{eq:19} depends, but for which the right hand side does not. For example, $\chi_{C}$ affects the value of $\xi_{CVR}$ on the left hand side but no quantity on the right hand side. If a parameter is changed such that

\begin{equation} \label{eq:20}
\xi_{CVR} > d \left(\frac{2 \pi m_{e} e^{2}}{\epsilon_{0}^{2} k^{3}}\right)^{1/4} \frac{J_{ES}^{1/2}}{T_{E}^{3/4}}
\end{equation}

\noindent then the device is operating within the ideal volume where it experiences no space charge limited mode. If the inequality is reversed, the device will pass through a space charge limited mode over some range of voltages.

\section{Conclusions}
A model was developed which considers the space charge limited electron transport through a TEC where the collector features NEA. Using this model, it is shown that NEA can mitigate, and in some cases eliminate, the negative space charge effect. Calculations were performed using the  model to show output characteristics of devices made form available materials. Using material parameters quoted in the literature, and making reasonable inferences on missing parameter values, the output power and efficiency of a plausible device was calculated to be $1.03 W cm^{-2}$ and $17.8\%$, respectively. Since the efficiency of the device was below the target of 20\%, the input parameters were individually adjusted. The device can reach the 20\% target when the collector barrier falls below $0.85eV$. At that point, the output power density is $1.17 W cm^{-2}$. Such a device would output $1kW$ with $0.086 m^{2}$ emission area. Increasing the emitter temperature to $1070K$ would also achieve 20\% efficiency.

\section{Colophon}
The model was implemented as a module written in python. Version 0.4.0 of the software implementing the model was used for the calculations in this manuscript along with scipy version 0.11.0, numpy version 1.7.0, matplotlib version 1.2.0, and python version 2.7.3 -- all installed using the Homebrew OS X package manager.  The code was executed on a MacBook Air model A1466 running OS X 10.8.4 (12E55), Darwin kernel version 12.4.0 with a 2GHz Intel core i7 and 8GB 1600 MHz DDR3 memory. The software to perform the calculations in this paper was developed according to the best practices advocated by the Software Carpentry project \cite{10.1109/MCSE.2006.122}.

\section{Acknowledgements}
Research was sponsored by the Army Research Laboratory and was accomplished under Cooperative Agreement Number W911NF-12-2-0019. The views and conclusions contained in this document are those of the authors and should not be interpreted as representing the official policies, either expressed or implied, of the Army Research Laboratory or the U.S. Government. The U.S. Government is authorized to reproduce and distribute reprints for Government purposes notwithstanding any copyright notation herein.

\bibliography{master.bib}

\begin{thebibliography}{18}%
\makeatletter
\providecommand \@ifxundefined [1]{%
 \@ifx{#1\undefined}
}%
\providecommand \@ifnum [1]{%
 \ifnum #1\expandafter \@firstoftwo
 \else \expandafter \@secondoftwo
 \fi
}%
\providecommand \@ifx [1]{%
 \ifx #1\expandafter \@firstoftwo
 \else \expandafter \@secondoftwo
 \fi
}%
\providecommand \natexlab [1]{#1}%
\providecommand \enquote  [1]{``#1''}%
\providecommand \bibnamefont  [1]{#1}%
\providecommand \bibfnamefont [1]{#1}%
\providecommand \citenamefont [1]{#1}%
\providecommand \href@noop [0]{\@secondoftwo}%
\providecommand \href [0]{\begingroup \@sanitize@url \@href}%
\providecommand \@href[1]{\@@startlink{#1}\@@href}%
\providecommand \@@href[1]{\endgroup#1\@@endlink}%
\providecommand \@sanitize@url [0]{\catcode `\\12\catcode `\$12\catcode
  `\&12\catcode `\#12\catcode `\^12\catcode `\_12\catcode `\%12\relax}%
\providecommand \@@startlink[1]{}%
\providecommand \@@endlink[0]{}%
\providecommand \url  [0]{\begingroup\@sanitize@url \@url }%
\providecommand \@url [1]{\endgroup\@href {#1}{\urlprefix }}%
\providecommand \urlprefix  [0]{URL }%
\providecommand \Eprint [0]{\href }%
\providecommand \doibase [0]{http://dx.doi.org/}%
\providecommand \selectlanguage [0]{\@gobble}%
\providecommand \bibinfo  [0]{\@secondoftwo}%
\providecommand \bibfield  [0]{\@secondoftwo}%
\providecommand \translation [1]{[#1]}%
\providecommand \BibitemOpen [0]{}%
\providecommand \bibitemStop [0]{}%
\providecommand \bibitemNoStop [0]{.\EOS\space}%
\providecommand \EOS [0]{\spacefactor3000\relax}%
\providecommand \BibitemShut  [1]{\csname bibitem#1\endcsname}%
\let\auto@bib@innerbib\@empty
\bibitem [{\citenamefont {{Committee on Thermionic Research and Technology,
  Aeronautics and Space Engineering Board, National Research
  Council}}(2001)}]{978-0-309-08686-8}%
  \BibitemOpen
  \bibfield  {author} {\bibinfo {author} {\bibnamefont {{Committee on
  Thermionic Research and Technology, Aeronautics and Space Engineering Board,
  National Research Council}}},\ }\href
  {http://www.nap.edu/openbook.php?record_id=10254} {\emph {\bibinfo {title}
  {Thermionics Quo Vadis? An Assessment of the DTRA's Advanced Thermionics
  Research and Development Program}}}\ (\bibinfo  {publisher} {The National
  Academies Press},\ \bibinfo {year} {2001})\BibitemShut {NoStop}%
\bibitem [{\citenamefont {Houston}(1959)}]{10.1063/1.1702392}%
  \BibitemOpen
  \bibfield  {author} {\bibinfo {author} {\bibfnamefont {J.~M.}\ \bibnamefont
  {Houston}},\ }\href {\doibase 10.1063/1.1702392} {\bibfield  {journal}
  {\bibinfo  {journal} {Journal of Applied Physics}\ }\textbf {\bibinfo
  {volume} {30}},\ \bibinfo {pages} {481} (\bibinfo {year} {1959})}\BibitemShut
  {NoStop}%
\bibitem [{\citenamefont {Koeck}\ and\ \citenamefont
  {Nemanich}(2009)}]{10.1016/j.diamond.2008.11.023}%
  \BibitemOpen
  \bibfield  {author} {\bibinfo {author} {\bibfnamefont {F.~A.}\ \bibnamefont
  {Koeck}}\ and\ \bibinfo {author} {\bibfnamefont {R.~J.}\ \bibnamefont
  {Nemanich}},\ }\href {\doibase 10.1016/j.diamond.2008.11.023} {\bibfield
  {journal} {\bibinfo  {journal} {Diamond and Related Materials}\ }\textbf
  {\bibinfo {volume} {18}},\ \bibinfo {pages} {232 } (\bibinfo {year}
  {2009})},\ \bibinfo {note} {nDNC 2008 Proceedings of the International
  Conference on New Diamond and Nano Carbons 2008}\BibitemShut {NoStop}%
\bibitem [{\citenamefont {Koeck}\ \emph {et~al.}(2009)\citenamefont {Koeck},
  \citenamefont {Nemanich}, \citenamefont {Lazea},\ and\ \citenamefont
  {Haenen}}]{10.1016/j.diamond.2009.01.024}%
  \BibitemOpen
  \bibfield  {author} {\bibinfo {author} {\bibfnamefont {F.~A.}\ \bibnamefont
  {Koeck}}, \bibinfo {author} {\bibfnamefont {R.~J.}\ \bibnamefont {Nemanich}},
  \bibinfo {author} {\bibfnamefont {A.}~\bibnamefont {Lazea}}, \ and\ \bibinfo
  {author} {\bibfnamefont {K.}~\bibnamefont {Haenen}},\ }\href {\doibase
  10.1016/j.diamond.2009.01.024} {\bibfield  {journal} {\bibinfo  {journal}
  {Diamond and Related Materials}\ }\textbf {\bibinfo {volume} {18}},\ \bibinfo
  {pages} {789 } (\bibinfo {year} {2009})},\ \bibinfo {note} {proceedings of
  Diamond 2008, the 19th European Conference on Diamond, Diamond-Like
  Materials, Carbon Nanotubes, Nitrides and Silicon Carbide}\BibitemShut
  {NoStop}%
\bibitem [{\citenamefont {Schwede}\ \emph {et~al.}(2010)\citenamefont
  {Schwede}, \citenamefont {Bargatin}, \citenamefont {Riley}, \citenamefont
  {Hardin}, \citenamefont {Rosenthal}, \citenamefont {Sun}, \citenamefont
  {Schmitt}, \citenamefont {Pianetta}, \citenamefont {Howe}, \citenamefont
  {Shen},\ and\ \citenamefont {Melosh}}]{10.1038/nmat2814}%
  \BibitemOpen
  \bibfield  {author} {\bibinfo {author} {\bibfnamefont {J.~W.}\ \bibnamefont
  {Schwede}}, \bibinfo {author} {\bibfnamefont {I.}~\bibnamefont {Bargatin}},
  \bibinfo {author} {\bibfnamefont {D.~C.}\ \bibnamefont {Riley}}, \bibinfo
  {author} {\bibfnamefont {B.~E.}\ \bibnamefont {Hardin}}, \bibinfo {author}
  {\bibfnamefont {S.~J.}\ \bibnamefont {Rosenthal}}, \bibinfo {author}
  {\bibfnamefont {Y.}~\bibnamefont {Sun}}, \bibinfo {author} {\bibfnamefont
  {F.}~\bibnamefont {Schmitt}}, \bibinfo {author} {\bibfnamefont
  {P.}~\bibnamefont {Pianetta}}, \bibinfo {author} {\bibfnamefont {R.~T.}\
  \bibnamefont {Howe}}, \bibinfo {author} {\bibfnamefont {Z.-X.}\ \bibnamefont
  {Shen}}, \ and\ \bibinfo {author} {\bibfnamefont {N.~A.}\ \bibnamefont
  {Melosh}},\ }\href {\doibase 10.1038/nmat2814} {\bibfield  {journal}
  {\bibinfo  {journal} {Nat Mater}\ }\textbf {\bibinfo {volume} {9}},\ \bibinfo
  {pages} {762} (\bibinfo {year} {2010})}\BibitemShut {NoStop}%
\bibitem [{\citenamefont {Sun}\ \emph {et~al.}(2011)\citenamefont {Sun},
  \citenamefont {Koeck}, \citenamefont {Zhu},\ and\ \citenamefont
  {Nemanich}}]{10.1063/1.3658638}%
  \BibitemOpen
  \bibfield  {author} {\bibinfo {author} {\bibfnamefont {T.}~\bibnamefont
  {Sun}}, \bibinfo {author} {\bibfnamefont {F.~A.~M.}\ \bibnamefont {Koeck}},
  \bibinfo {author} {\bibfnamefont {C.}~\bibnamefont {Zhu}}, \ and\ \bibinfo
  {author} {\bibfnamefont {R.~J.}\ \bibnamefont {Nemanich}},\ }\href {\doibase
  10.1063/1.3658638} {\bibfield  {journal} {\bibinfo  {journal} {Applied
  Physics Letters}\ }\textbf {\bibinfo {volume} {99}},\ \bibinfo {eid} {202101}
  (\bibinfo {year} {2011})}\BibitemShut {NoStop}%
\bibitem [{\citenamefont {Hatsopoulos}\ and\ \citenamefont
  {Kaye}(1958)}]{10.1063/1.1723373}%
  \BibitemOpen
  \bibfield  {author} {\bibinfo {author} {\bibfnamefont {G.~N.}\ \bibnamefont
  {Hatsopoulos}}\ and\ \bibinfo {author} {\bibfnamefont {J.}~\bibnamefont
  {Kaye}},\ }\href {\doibase 10.1063/1.1723373} {\bibfield  {journal} {\bibinfo
   {journal} {Journal of Applied Physics}\ }\textbf {\bibinfo {volume} {29}},\
  \bibinfo {pages} {1124} (\bibinfo {year} {1958})}\BibitemShut {NoStop}%
\bibitem [{\citenamefont {Lee}\ \emph {et~al.}(2012{\natexlab{a}})\citenamefont
  {Lee}, \citenamefont {Bargatin}, \citenamefont {Iwami}, \citenamefont
  {Littau}, \citenamefont {Vincent}, \citenamefont {Maboudian}, \citenamefont
  {Shen}, \citenamefont {Melosh},\ and\ \citenamefont {Howe}}]{lee}%
  \BibitemOpen
  \bibfield  {author} {\bibinfo {author} {\bibfnamefont {J.}~\bibnamefont
  {Lee}}, \bibinfo {author} {\bibfnamefont {I.}~\bibnamefont {Bargatin}},
  \bibinfo {author} {\bibfnamefont {K.}~\bibnamefont {Iwami}}, \bibinfo
  {author} {\bibfnamefont {K.}~\bibnamefont {Littau}}, \bibinfo {author}
  {\bibfnamefont {M.}~\bibnamefont {Vincent}}, \bibinfo {author} {\bibfnamefont
  {R.}~\bibnamefont {Maboudian}}, \bibinfo {author} {\bibfnamefont {Z.-X.}\
  \bibnamefont {Shen}}, \bibinfo {author} {\bibfnamefont {N.}~\bibnamefont
  {Melosh}}, \ and\ \bibinfo {author} {\bibfnamefont {R.}~\bibnamefont
  {Howe}},\ }in\ \href@noop {} {\emph {\bibinfo {booktitle} {Workshop on
  Solid-state Sensors, Actuators, and Microsystems Workshop}}}\ (\bibinfo
  {address} {Hilton Head, SC},\ \bibinfo {year} {2012})\BibitemShut {NoStop}%
\bibitem [{\citenamefont {Smith}\ \emph {et~al.}(2007)\citenamefont {Smith},
  \citenamefont {Bilbro},\ and\ \citenamefont
  {Nemanich}}]{10.1103/PhysRevB.76.245327}%
  \BibitemOpen
  \bibfield  {author} {\bibinfo {author} {\bibfnamefont {J.~R.}\ \bibnamefont
  {Smith}}, \bibinfo {author} {\bibfnamefont {G.~L.}\ \bibnamefont {Bilbro}}, \
  and\ \bibinfo {author} {\bibfnamefont {R.~J.}\ \bibnamefont {Nemanich}},\
  }\href {\doibase 10.1103/PhysRevB.76.245327} {\bibfield  {journal} {\bibinfo
  {journal} {Phys. Rev. B}\ }\textbf {\bibinfo {volume} {76}},\ \bibinfo
  {pages} {245327} (\bibinfo {year} {2007})}\BibitemShut {NoStop}%
\bibitem [{\citenamefont {Smith}\ \emph {et~al.}(2009)\citenamefont {Smith},
  \citenamefont {Bilbro},\ and\ \citenamefont {Nemanich}}]{10.1116/1.3125282}%
  \BibitemOpen
  \bibfield  {author} {\bibinfo {author} {\bibfnamefont {J.~R.}\ \bibnamefont
  {Smith}}, \bibinfo {author} {\bibfnamefont {G.~L.}\ \bibnamefont {Bilbro}}, \
  and\ \bibinfo {author} {\bibfnamefont {R.~J.}\ \bibnamefont {Nemanich}},\
  }\href {\doibase 10.1116/1.3125282} {\bibfield  {journal} {\bibinfo
  {journal} {Journal of Vacuum Science \& Technology B: Microelectronics and
  Nanometer Structures}\ }\textbf {\bibinfo {volume} {27}},\ \bibinfo {pages}
  {1132} (\bibinfo {year} {2009})}\BibitemShut {NoStop}%
\bibitem [{\citenamefont {G\"{a}rtner}\ \emph {et~al.}(1997)\citenamefont
  {G\"{a}rtner}, \citenamefont {Geittner}, \citenamefont {Lydtin},\ and\
  \citenamefont {Ritz}}]{10.1016/S0169-4332(96)00698-8}%
  \BibitemOpen
  \bibfield  {author} {\bibinfo {author} {\bibfnamefont {G.}~\bibnamefont
  {G\"{a}rtner}}, \bibinfo {author} {\bibfnamefont {P.}~\bibnamefont
  {Geittner}}, \bibinfo {author} {\bibfnamefont {H.}~\bibnamefont {Lydtin}}, \
  and\ \bibinfo {author} {\bibfnamefont {A.}~\bibnamefont {Ritz}},\ }\href
  {\doibase 10.1016/S0169-4332(96)00698-8} {\bibfield  {journal} {\bibinfo
  {journal} {Applied Surface Science}\ }\textbf {\bibinfo {volume} {111}},\
  \bibinfo {pages} {11 } (\bibinfo {year} {1997})},\ \bibinfo {note}
  {proceedings of the International Vacuum Electron Sources Conference
  1996}\BibitemShut {NoStop}%
\bibitem [{\citenamefont {Langmuir}(1923)}]{10.1103/PhysRev.21.419}%
  \BibitemOpen
  \bibfield  {author} {\bibinfo {author} {\bibfnamefont {I.}~\bibnamefont
  {Langmuir}},\ }\href {\doibase 10.1103/PhysRev.21.419} {\bibfield  {journal}
  {\bibinfo  {journal} {Phys. Rev.}\ }\textbf {\bibinfo {volume} {21}},\
  \bibinfo {pages} {419} (\bibinfo {year} {1923})}\BibitemShut {NoStop}%
\bibitem [{\citenamefont {Dugan}(1960)}]{10.1063/1.1735850}%
  \BibitemOpen
  \bibfield  {author} {\bibinfo {author} {\bibfnamefont {A.~F.}\ \bibnamefont
  {Dugan}},\ }\href {\doibase 10.1063/1.1735850} {\bibfield  {journal}
  {\bibinfo  {journal} {Journal of Applied Physics}\ }\textbf {\bibinfo
  {volume} {31}},\ \bibinfo {pages} {1397} (\bibinfo {year}
  {1960})}\BibitemShut {NoStop}%
\bibitem [{\citenamefont {Hatsopoulous}\ and\ \citenamefont
  {Gyftopoulos}(1973)}]{0-262-08059-1}%
  \BibitemOpen
  \bibfield  {author} {\bibinfo {author} {\bibfnamefont {G.}~\bibnamefont
  {Hatsopoulous}}\ and\ \bibinfo {author} {\bibfnamefont {E.}~\bibnamefont
  {Gyftopoulos}},\ }\href@noop {} {\emph {\bibinfo {title} {Thermionic Energy
  Conversion}}},\ Vol.~\bibinfo {volume} {1}\ (\bibinfo  {publisher} {MIT
  Press},\ \bibinfo {address} {Cambridge, Massachusetts},\ \bibinfo {year}
  {1973})\ p.\ \bibinfo {pages} {265}\BibitemShut {NoStop}%
\bibitem [{\citenamefont {van~der Weide}\ \emph {et~al.}(1994)\citenamefont
  {van~der Weide}, \citenamefont {Zhang}, \citenamefont {Baumann},
  \citenamefont {Wensell}, \citenamefont {Bernholc},\ and\ \citenamefont
  {Nemanich}}]{10.1103/PhysRevB.50.5803}%
  \BibitemOpen
  \bibfield  {author} {\bibinfo {author} {\bibfnamefont {J.}~\bibnamefont
  {van~der Weide}}, \bibinfo {author} {\bibfnamefont {Z.}~\bibnamefont
  {Zhang}}, \bibinfo {author} {\bibfnamefont {P.~K.}\ \bibnamefont {Baumann}},
  \bibinfo {author} {\bibfnamefont {M.~G.}\ \bibnamefont {Wensell}}, \bibinfo
  {author} {\bibfnamefont {J.}~\bibnamefont {Bernholc}}, \ and\ \bibinfo
  {author} {\bibfnamefont {R.~J.}\ \bibnamefont {Nemanich}},\ }\href {\doibase
  10.1103/PhysRevB.50.5803} {\bibfield  {journal} {\bibinfo  {journal} {Phys.
  Rev. B}\ }\textbf {\bibinfo {volume} {50}},\ \bibinfo {pages} {5803}
  (\bibinfo {year} {1994})}\BibitemShut {NoStop}%
\bibitem [{\citenamefont {Cui}\ \emph {et~al.}(1998)\citenamefont {Cui},
  \citenamefont {Ristein},\ and\ \citenamefont
  {Ley}}]{10.1103/PhysRevLett.81.429}%
  \BibitemOpen
  \bibfield  {author} {\bibinfo {author} {\bibfnamefont {J.~B.}\ \bibnamefont
  {Cui}}, \bibinfo {author} {\bibfnamefont {J.}~\bibnamefont {Ristein}}, \ and\
  \bibinfo {author} {\bibfnamefont {L.}~\bibnamefont {Ley}},\ }\href {\doibase
  10.1103/PhysRevLett.81.429} {\bibfield  {journal} {\bibinfo  {journal} {Phys.
  Rev. Lett.}\ }\textbf {\bibinfo {volume} {81}},\ \bibinfo {pages} {429}
  (\bibinfo {year} {1998})}\BibitemShut {NoStop}%
\bibitem [{\citenamefont {Lee}\ \emph {et~al.}(2012{\natexlab{b}})\citenamefont
  {Lee}, \citenamefont {Bargatin}, \citenamefont {Melosh},\ and\ \citenamefont
  {Howe}}]{10.1063/1.4707379}%
  \BibitemOpen
  \bibfield  {author} {\bibinfo {author} {\bibfnamefont {J.-H.}\ \bibnamefont
  {Lee}}, \bibinfo {author} {\bibfnamefont {I.}~\bibnamefont {Bargatin}},
  \bibinfo {author} {\bibfnamefont {N.~A.}\ \bibnamefont {Melosh}}, \ and\
  \bibinfo {author} {\bibfnamefont {R.~T.}\ \bibnamefont {Howe}},\ }\href
  {\doibase 10.1063/1.4707379} {\bibfield  {journal} {\bibinfo  {journal}
  {Applied Physics Letters}\ }\textbf {\bibinfo {volume} {100}},\ \bibinfo
  {eid} {173904} (\bibinfo {year} {2012}{\natexlab{b}})}\BibitemShut {NoStop}%
\bibitem [{\citenamefont {Wilson}(2006)}]{10.1109/MCSE.2006.122}%
  \BibitemOpen
  \bibfield  {author} {\bibinfo {author} {\bibfnamefont {G.}~\bibnamefont
  {Wilson}},\ }\href {\doibase 10.1109/MCSE.2006.122} {\bibfield  {journal}
  {\bibinfo  {journal} {Computing in Science \& Engineering}\ }\textbf
  {\bibinfo {volume} {8}},\ \bibinfo {pages} {66} (\bibinfo {year}
  {2006})}\BibitemShut {NoStop}%
\end{thebibliography}%

\end{document}